# Development of fall prevention training device that can provide external disturbance to the ankle with pneumatic gel muscles (PGM) while walking

Keigo Isoshima[1], Mitsunori Tada[2], Noriaki Maeda[3], Tsubasa Tashiro[3], Satoshi Arima[3], Takumi Nagao[3], Yuki Tamura[3], and Yuichi Kurita[1]

[1] *Graduate School of Advanced Science and Engineering, Hiroshima University, Higashi Hiroshima, Japan*
[2] *Digital Human Research Team, Artificial Intelligence Research Center (AIST), Tokyo, Japan*
[3] *Graduate School of Biomedical and Health Sciences, Hiroshima University, Hiroshima, Japan*
(Email: keigoisoshima@hiroshima-u.ac.jp)

**Abstract ---** Although the average life expectancy in Japan has been increasing in recent years, the problem of the large gap between healthy and average life expectancy still needs to be solved. Among the factors that lead to the need for nursing care, injuries due to falls account for a certain percentage of the total. In this paper, we developed boots that can provide external disturbance to the ankle with pneumatic gel muscles (PGM) while walking. We experimented using the acceleration and angular velocity of the heel as evaluation indices to evaluate the effectiveness of fall prevention training using this device, which is smaller and more wearable than conventional devices. This study confirmed that the developed system has enough training intensity to affect the gait waveform significantly.

## 1 INTRODUCTION

In fall prevention, training is commonly provided through therapist intervention. One conventional example of such training is a patient suspended from the ceiling and pushed by a therapist to withstand the disturbance [1]. However, because another person provides the intervention, it cannot be said that the caregiver can perform the training spontaneously. In addition, the therapist cannot leave the room during training, making the conventional method burdensome for both the patient and the therapist. Against this background, research has been conducted on disturbance training using a balance suit worn on the upper body [2] and disturbance training on the lower body using a moving platform as fall prevention training [3] that does not require intervention from others. The results of these studies have shown that disturbance training improves the indexes' effectiveness in fall prevention. However, the former only considered upper-body interventions, while the latter required extensive equipment. Therefore, we integrated the findings from these studies and aimed to realize a new fall prevention training method that uses a small, variable-ability brace that can conform to the wearer's body shape to provide disturbance to the lower body.

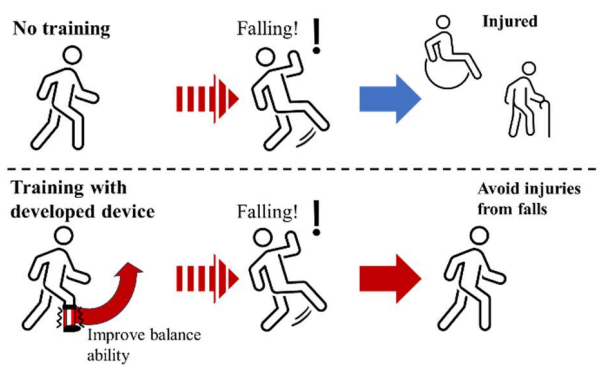

Fig.1 Conceptual diagram of this study

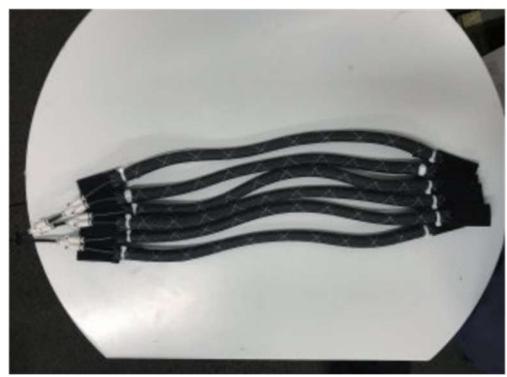

Fig.2 Pneumatic Gel Muscle

Fig.1 shows a conceptual diagram of this study. In a previous study, we examined whether a small orthosis powered by low-pressure-driven pneumatic artificial muscles（Pneumatic Gel Muscle： PGM[4]）( Fig.2 ), which can exert a contractile force by applying pressure, can bring about a practical fall prevention effect on the human body through minute disturbance to the lower body [5]. As a result, we succeeded in clarifying that a small device using PGM can significantly affect the balance control of the human body during one-leg standing training. However, there was an issue with the fall prevention training in this previous study, which remained static, unlike actual fall situations. In terms of actual fall situations, it has been shown that the majority of falls occur while walking [6]. Regarding training due to disturbances during walking, dynamic training using a dual treadmill [7] has been reported to improve the ability to control the center of pressure (COP), which suggests the importance of dynamic training.

In this study, we developed a new system that applies a disturbance to the lower body while walking, not while standing still, using a PGM-powered orthosis. We also conducted a walking task experiment on healthy participants to verify the feasibility of dynamic training, which had not been tested in previous studies.

## 2 METHOD

### 2.1 System Configurations

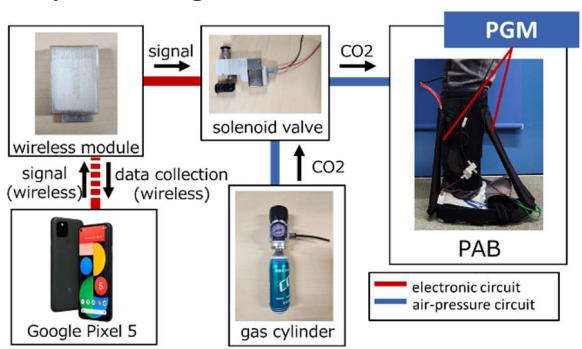

Fig.3  Overall system configuration

The following is an overview of the constructed disturbance application system. The system is based on an improved assist boot (DARWING Power Assist Boots by Dia Industry, hereinafter referred to as "PAB"), a PGM-powered walking assist device. It is controlled by a small wireless module (DhaibaDAQ) capable of estimating the phase of gait. The overall system configuration is shown in Fig.3.

### 2.2 Disturbance-giving Part

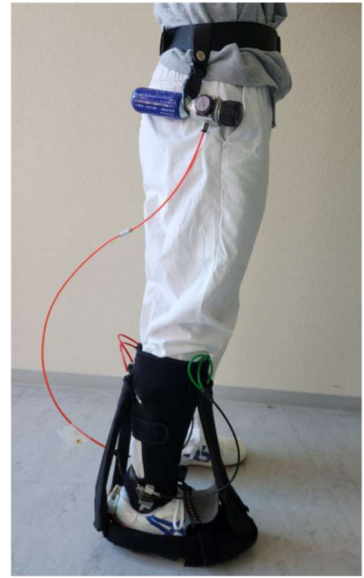

Fig.4  Developed device

The PAB is equipped with four PGMs, two in the front and two in the rear, which contract when supplied with air, allowing the front two to contract to generate a forward disturbance and the rear two to contract to generate a backward disturbance. As shown in Fig.4  the developed device is compact and easy to wear. The total weight of the device is about 1 [kg], which is lightweight and addresses the problem of large equipment that has been a problem with conventional devices.

### 2.3 Experimental Method

Ten healthy participants (10 males) aged 21-23 years were asked to perform a walking task of approximately 10[m] several times, and their heels' triaxial acceleration and angular velocity were measured. A small wireless module attached to the heel was used for measurement.

The task was performed in three phases: (1) walking without intervention, (2) walking with intervention, and (3) walking after intervention. In all phases, the participants wore the developed device. Considering the risk of falling, one researcher watched the subjects closely while they performed all task movements and took sufficient care to prevent them from falling. All subjects wore the device only on the designated leg. The details of each task are described below.

Table 1 shows a summary of the experimental conditions. Walking without intervention and walking after intervention were tasks to record normal walking before and after the intervention by the developed device to be compared in the analysis. Therefore, each task was performed only once, without any intervention by the device. The intervention gait was a task to confirm the training intensity of the device. Subjects walked while wearing the device, and a disturbance of 0.5 [s] was applied randomly. The timing of the disturbance was based on the gait phase calculated from the gait cycle estimated by the wireless module.

The human gait is divided into 8 phases [Initial Contact (IC), Loading Response (LR), Mid Stance (MSt), Terminal Stance (TSt), Pre-Swing (PSw), Initial Swing (ISw), Mid Swing (MSw), and Terminal Swing (TSw)], and the percentage of time spent in each gait phase is approximately IC(0%), LR (0%~12%), MSt (12%~31%), TSt (31%~50%), PSw (50%~62%), ISw (62%~75%), MSw (75%~87%), and TSw (87%~100%) when the gait cycle was set to 100% [8]. In this experiment, the gait cycle was divided into four phases, IC~LR, MSt~TSt, PSw~ISw, and MSw~TSw, and the PGM was controlled so that it was randomly selected in which phase it was driven and in which direction it was driven forward or backward. In all tasks, subjects were instructed to walk in a steady rhythm as much as possible.

This experiment was conducted under the approval of the Research Ethics Review Committee of the Graduate School of Advanced Science and Engineering, Hiroshima University (Approval No.: ASE-2023-21).

Table 1  Experimental conditions

| Task Name | Disturbance | Task Count |
|---|---|---|
| walking without intervention | No Disturbance | 1 |
| walking with intervention | Apply Disturbance | 5 |
| walking after intervention | No Disturbance | 1 |

**2.4 Analysis Method**

First, the three axial acceleration and angular velocity waveforms obtained during the initial walking task (walking without intervention) were separated into step-by-step data. Next, the data of several steps obtained in the initial walking task (walking without intervention) were resampled by linear completion to align their data widths and averaged to create a single initial walking waveform for each subject. The initial gait waveform obtained here is used as the basis for all subsequent calculations.

After obtaining the reference waveform as described above, the data for the intervention gait (walking with intervention) and post-intervention gait (walking after intervention) were similarly segmented, and the similarity was calculated by comparing the segmented waveforms with the reference waveform. The similarity was calculated using Dynamic Time Warping (DTW), which can calculate the distance and similarity considering the shape of the upward and downward trends of the time series data.

In the following sections, the similarity of the waveforms will be evaluated based on the numerical value of the distance calculated by DTW and will be used as an indicator of whether or not the developed device influenced the subject's gait.

**3 RESULTS AND DISCUSSION**

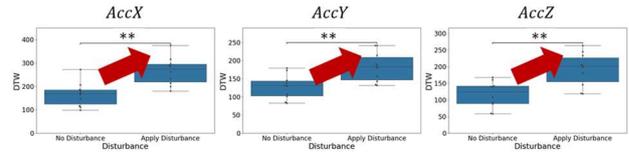

Fig.5  Experimental results of the DTW (Acceleration)

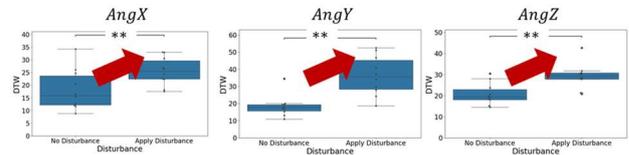

Fig.6  Experimental results of the DTW (Angular velocity)

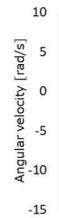

Fig.7  Waveform of AngY (No Disturbance)

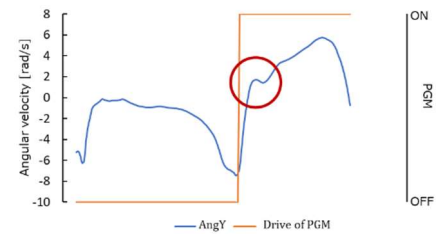

Fig.8  Waveform of AngY (Apply Disturbance)

The results of the analysis for acceleration ($AccX, AccY, AccZ$) and angular velocity ($AngX, AngY, AngZ$) in the X, Y, and Z axes are shown in Fig.5 and Fig.6 They show the DTW calculated between the waveform of the initial gait and the walk without intervention (No Disturbance), and between the waveform of the initial gait and the walk with intervention (Disturbance) and the results of the significance tests for each acceleration and angular velocity. A Wilcoxon signed-rank test was used for the test between the two groups corresponding to each measurement, and the significance level was set at 5% (**: p<0.01). From there, it can be confirmed that the disturbance by the developed device significantly changed the gait waveform from the initial gait, and it can be said that the disturbance in this experiment succeeded in disrupting the gait waveform for healthy subjects.

It is known that the angular velocity waveform is almost constant regardless of walking speed [9]. Therefore, the angular velocity results on the y-axis for one subject are shown below, with time on the horizontal axis and angular velocity on the vertical axis. Fig.7 shows the calculated initial gait waveform, and Fig.8 shows the waveform measured during one of the intervention tasks. Comparing Fig.7 and Fig.8 it can be seen that the angular velocity waveform is deformed from the initial walking waveform immediately after PGM drives (the area indicated by the red circle). From these results, it can be said that the small device using PGM may have a certain level of training intensity.

## 4 Conclusion

In this study, we developed a system that enables fall prevention training by applying disturbance to the ankle during walking and verified the effect of the system by analyzing the similarity of the gait waveforms of healthy university students during a walking task using the system. The results confirmed that the developed system has enough training intensity to affect the gait waveform significantly. However, the number of subjects was small, and the experiment was conducted on healthy university students, so the improvement of walking ability before and after the gait task could not be examined. In the future, we will increase the number of subjects and conduct a more detailed analysis to examine further the possibility of improving motor function.

Acknowledgment

This work was supported by JSPS KAKENHI Grant Numbers 21H01292, 22K18418 and 22H00526.